\documentclass{elsart3p}

\usepackage{graphicx}
\usepackage{lineno}
\usepackage{amssymb}
\usepackage{epsfig}
\usepackage{amsmath}
\usepackage{xspace}

\usepackage{ifpdf}

\begin{document}
\begin{frontmatter}

\title{An alpha tagged X-ray source for the calibration of space
borne X-ray detectors 
         \thanksref{cor}}
\thanks[cor]{Send correspondence to A. R. Rao: \\
        E-mail: arrao@tifr.res.in}
 
\author{A. R. Rao}
\address{Department of Astronomy and Astrophysics, Tata Institute of Fundamental Research, Homi Bhabha road, Mumbai, India}
\author{Sachindra Naik}
\address{Astronomy and Astrophysics Division, Physical Research Laboratory, Navrangpura, Ahmedabad - 380009, India}
\author{Milind Patil, J. P. Malkar}
\address{Department of Astronomy and Astrophysics, Tata Institute of Fundamental Research, Homi Bhabha road, Mumbai, India}
\author{R. P. S. Kalyan Kumar}
\address{Department of Physics, Pondicherry Univeristy, Puducherry, India}

\begin{abstract}
	Calibration of X-ray detectors is very important to understand
the performance characteristics of the detectors and their variation with 
time and changing operational conditions. This  enables the most accurate 
translation of the measurements to absolute and relative values of the 
incident X-ray photon energy so that physical models of the source emission
can be tested. It is a general practice to put a known X-ray source (radio
active source) in the detector housing for the calibration purpose. This,
however, increases the background. Tagging the calibration source with the
signal from a simultaneously emitted charge particle (like alpha particle) 
can identify the X-ray event used for calibration. Here in this paper, we
present a new design for an   alpha-tagged X-ray source using Am$^{241}$ 
radio active source and describe its performance characteristics. Its
application for the upcoming Astrosat satellite is also discussed.
\end{abstract}


\end{frontmatter}

\section{Introduction}

\begin{figure*}
\centering
\includegraphics*[width=9cm,angle=-90]{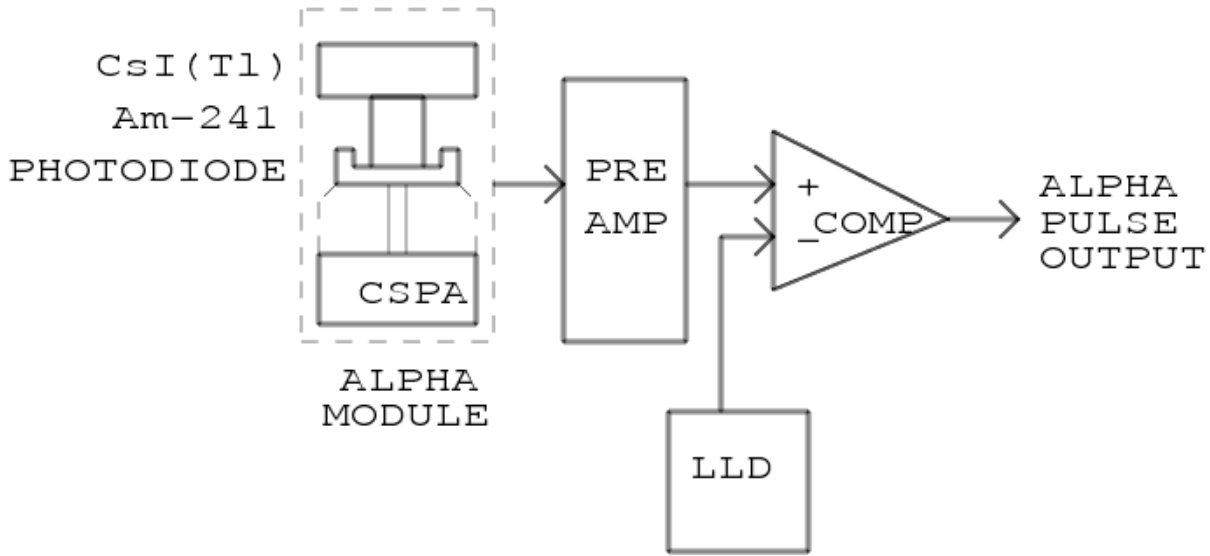}
\caption{The block diagram of Alpha-tag source with the Alpha Module
and the electronics for generating Alpha pulse output for tagging the 
X-rays.}
\label{fig1}
\end{figure*}

The characterization and measurement of X-ray emission from the astrophysical
sources require well calibrated X-ray detectors. Parameterizing the energy
scale and monitoring the variations in the performance of X-ray detectors 
can be done by using an internal calibration source which emits continuous
calibration lines at certain energies. The overall performance can further 
be verified by analyzing the data recorded from the observations of any 
standard X-ray source in the sky. One commonly used radio active source 
is Americium-241 (Am$^{241}$), which  provides a continuous source of
calibration lines with energies between 13 and 60 keV. For example,
the calibration of the Proportional Counter Arrays (PCA) onboard the
Rossi X-ray Timing Explorer (RXTE) is done using a  Am$^{241}$ source
embedded in a proportional counter [1,2]. Similarly, a tagged Am$^{241}$ 
source is used to calibrate and stabilize the gain in the High Energy 
X-ray Timing Explorer (HEXTE) of the RXTE satellite [3,4]. The 
calibration source establishes the energy scale and its variation
with time and other parameters along with the relative variation of
the efficiency. The broadband performance and the flux calibration, 
however, is usually done by observing a few standard X-ray sources in 
sky such as Crab Nebula  which provide a bright stable broadband source 
with a well-characterized spectrum.
 
Americium is a human-made radio active element with atomic number of 95.
Americium-241 (Am$^{241}$) can be produced by bombarding plutonium-234 
with alpha particles. Am$^{241}$, with a half-life of 432.2 years, decays
primarily by alpha particle emission (at 5485 keV - 84.5\% and 5443 keV - 
13\%) to Neptunium-237 which has  a half-life of 2.144$\times$10$^{6}$ 
years. These decays are accompanied by low energy gamma radiation with 
the 59.5 keV gamma emission (35.9\%) being the most important one along 
with 26.35 keV (2.4\%) and 13.9 keV (42\%) emission. The radio active 
source Am$^{241}$ is commonly used for testing and/or calibrating 
the majority of hard X-ray detectors specifically because of the presence 
of the prominent 59.54 keV X-ray emission.
 
To detect the alpha particles simultaneous to the X-ray emission, the 
Am$^{241}$ source is embedded in a radiation detector which detects the 
alpha particles but is transparent to the X-rays. It can be done by using
proportional counters (as was done for the PCA detectors [2]) or thin
scintillators viewed by a Photo-multiplier tube (PMT), as was done
for HEXTE [3]. Both these methods, however, require the generation and 
use of high voltages (a few hundred volts) which is cumbersome and heavy. 
For the Astrosat satellite [5], a large area hard X-ray detector of area 
1000 cm$^2$ using an array of Cadmium Zinc Telluride (CZT) detectors is 
being developed with a overall mass budget of 50 kg. To calibrate these
detectors (called CZT-Imager or CZTI), we have developed a compact low 
mass alpha-tagged X-ray source. 

In this paper, we present the design of a new method of using Am$^{241}$
as an alpha-tagged X-ray source and discuss its efficiency as a
calibration source for the CZTI of the Astrosat satellite.

\section{Design of the alpha-tag X-ray source }

We have designed a very simple compact system for detecting the 
alpha particle coming from the Am$^{241}$ radio active source. In 
addition to the detection of alpha particles, the system also transmits 
the 59.54 keV X-ray photons emitted by Am$^{241}$ which can be used for 
the purpose of calibration of any radiation detector. In this detector 
system, the principal components used are (i) the radio active source 
Am$^{241}$ which emits the alpha particles as well as the 59.54 keV 
X-rays, (ii) a CsI(TI) crystal which detects the alpha particles and 
also transmits the 59.54 keV X-rays, (iii) a Si-PIN photo-diode for the 
detection of the light output from the CsI(TI) crystal (due to the 
interaction of the alpha particle) iv) a charge sensitive pre-amplifier
(CSPA) mounted right behind the  photo-diode. This whole system, called 
the Alpha Module, is light weight ($\sim$3 gm) and requires only 
a low voltage supply (+12 V) with a power dissipation of 130 mW.

\begin{figure}
\centering
\includegraphics*[width=5.5cm,angle=-90]{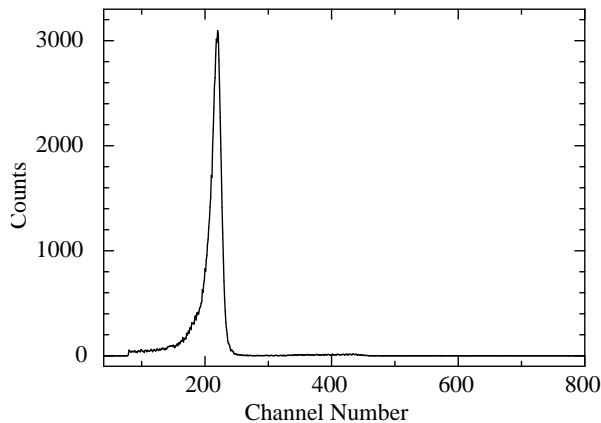}
\caption{The pulse height distribution from the Alpha-tag source. 
The peak corresponds to the 5.5 MeV alpha particles.}
\label{fig2}
\end{figure}

\begin{figure*}
\centering
\includegraphics*[width=10cm,angle=-90]{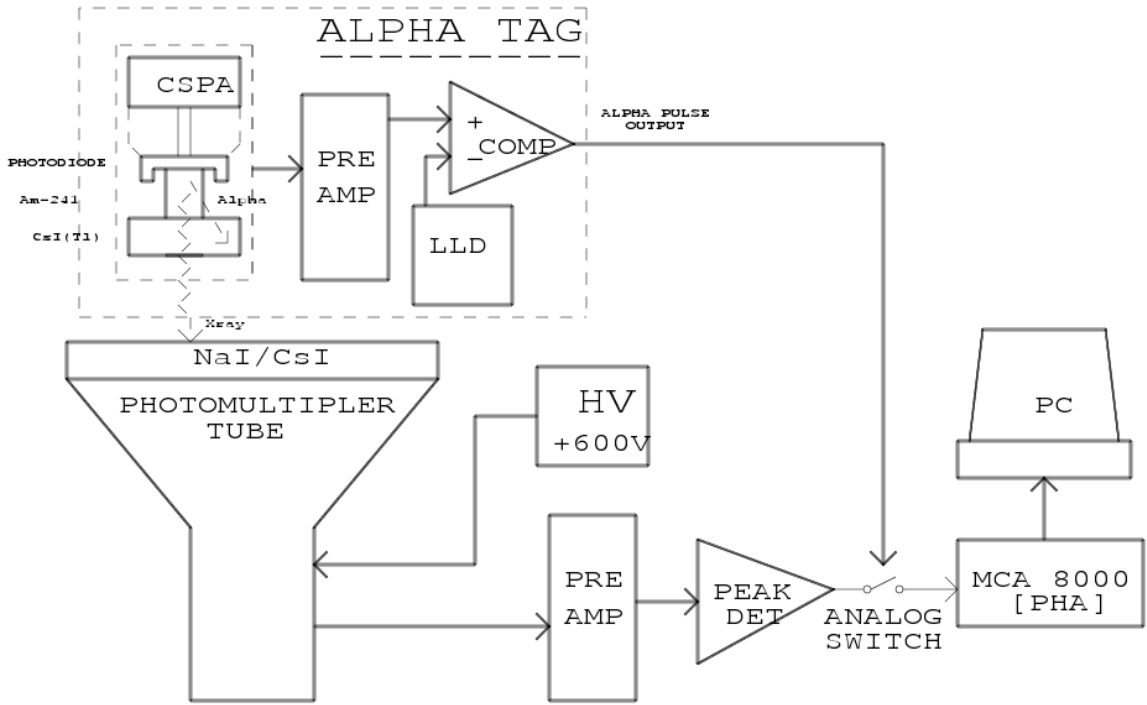}
\caption{The experimental set-up to tests the alpha tagging efficiency of the
the Alpha-tag source.}
\label{fig3}
\end{figure*}

The Alpha Module (see Figure~\ref{fig1}) is a small 10$\times$10$\times$0.3 
mm$^3$ of CsI (Tl) crystal (wavelength = 550 $nm$) with Teflon reflective
material coupled to a Si-PIN photo-diode of the same area. It contains a 
small pallet of Am$^{241}$ with a low strength of about 2000 Bq which is 
mounted on a 0.15 mm steel template. This radioactive source is kept in 
contact with a 0.3 mm CsI(Tl) crystal that detects the alpha particles and 
also transmits the 59.54 keV X-rays with a transparency of 35\%. The Si-PIN
photo-diode  detects the light output from the CsI(Tl) crystal (due to the
interaction of the alpha particle). This photo-diode has a broadband 
response in visible spectrum (average efficiency of 50\%) with higher
sensitivity and low dark current. Incident energy from the alpha particle 
is converted into light in CsI(Tl) whereas the light seen by the 
photo-diode is converted into an electrical pulse, which is amplified by 
the charge sensitive pre-amplifier (CSPA). The CsI(Tl) along with 
photo-diode and CSPA, as a whole structure, is called  the Alpha Module, 
which is rugged, compact, reliable and does not need a high voltage supply 
for biasing purposes. This whole compact system is procured from M/s Scionix,
Holand. Since the photo-diode output impedance is high (10 Mega ohms) and 
signal level is of the order of a few milli volts, a low noise CSPA is used 
to process the signal. This CSPA is placed just behind the photo-diode, so 
as to reduce noise. The pre-amplifier is wired on a 10$\times$10 mm$^2$ 
PCB (printed circuit board) and it is an integral part of the Alpha Module.

We have designed a simple electronics circuit to read out the pulses from
the Alpha Module. The signal from the CSPA is passed through a pre-amplifier
circuit with appropriate integration to achieve bipolarity with a gain of 
900. The sharp pulses (rise time ~0.5 $\mu$s) resulting from the detection 
of ionizing radiation  (signal from the CsI(Tl) crystal through the 
photo-diode) are processed in the amplifier to have suitable shape and
amplification. This amplified signal is passed through a comparator having 
a fixed LLD (lower-level discriminator) bias to cut off noise. This whole 
scheme is shown in Figure 1, along with the Alpha Module. The module can be 
used as a source of X-rays from Am$^{241}$ and the Alpha pulse output from 
this device can be used as a tag for the X-rays. The Alpha Module, mounted 
on a PCB, along with the electronics, is called the Alpha-tag source and it 
is used in the CZTI payload of Astrosat for in-flight calibration.

The amplifier output is fed to a Multi Channel Analyzer (MCA) and the 
results are shown in Figure~\ref{fig2}. A peak corresponding to the 5.5 MeV
alpha particles is clearly seen in the figure. A count rate corresponding to 
720 counts per second is noted.

\section{X-rays from the Alpha-tag source and the alpha-tagging methodology }

We made an experimental set up to study the alpha-tagged X-ray source
characteristics. Our experiment consists of a NaI(Tl)/CsI(Na) Phoswich
scintillator detector on the top of which the Alpha-tag source was mounted.
The  scintillators are coupled to a  PMT which multiplies  the current 
produced by the incident light by as much as 100 million times (i.e. 160 dB), 
in multiple dynode stages, enabling individual photons to be detected. The
current signal from PMT for every incident energy is converted to voltage 
and then passed  through an amplifier having a gain of around 20. 

The alpha particles from the Americium source in the Alpha Module generates
pulses in the Alpha-tag source electronics (the circuit diagram of which is
shown in Figure~1 along with the Alpha Module) and the 59.54 keV X-rays, if
they interact with the Phoswich detector, generates signal in the PMT 
electronics. The signal from PMT was then gated with the Alpha pulse and 
plotted on computer using a pulse height analyzer (PHA) software. It was 
found that there was a delay of approximately 5 to  15 $\mu$s in the Alpha 
pulse with respect to the X-ray pulse, due to the different electronics
characteristics. This delay was accommodated by adding a peak detector 
circuit in the path of the PMT signal. The  amplified signal was then 
connected directly to a MCA and/or an oscilloscope. The pocket MCA (MCA 
8000A from AMPTEK) was used in the experiment, which was again connected 
to a computer for data acquisition and storage. The experimental set-up is 
shown in Figure~\ref{fig3}. 

\begin{figure}
\centering
\includegraphics*[width=5.5cm,angle=-90]{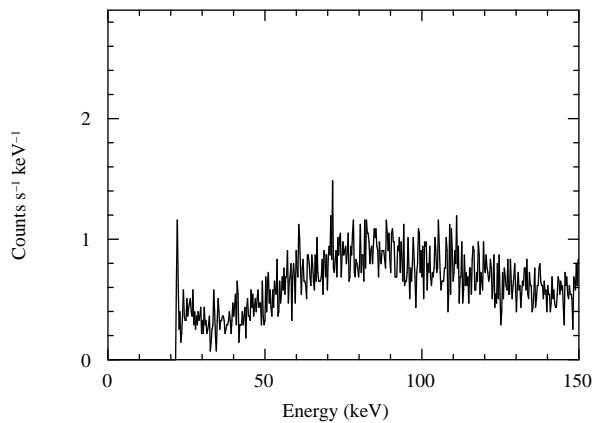}
\caption{The background spectrum from the Phoswich detector.}
\label{fig4}
\end{figure}

\begin{figure}
\centering
\includegraphics*[width=5.5cm,angle=-90]{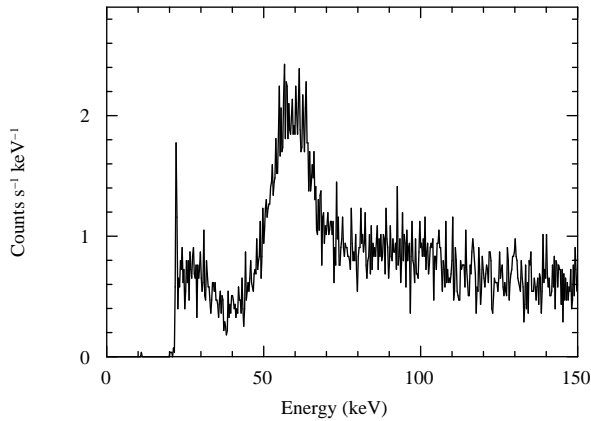}
\caption{The spectrum from the Phoswich detector with the alpha-source 
kept on it.}
\label{fig5}
\end{figure}

To reduce external noise, necessary care was taken to block extraneous 
light entering the PMT. A thick lead sheet with a small hole at the center 
was used as a collimator on the top of the Phoswich detector to block 
X-rays coming from other directions to reduce the  background. An 
arrangement was also made to block the X-rays from the Alpha-tag source
so that background could be measured from the Phoswich detectors without 
actually removing the  Alpha-tag source. With this set-up, the data were recorded by blocking the X-rays from the alpha source to measure the ambient
background. The MCA was calibrated by shining a bright Am$^{241}$ source 
directly onto the Phoswich detector and measuring the peak corresponding 
to the 60 keV X-rays. The Alpha-tag source was then kept on the Phoswich 
set-up and the spectrum  was taken by repeating the above procedure. Excess
counts were seen near the 60 keV peak and we note a count rate of  23 
counts/s due to the Am$^{241}$ source. The 2000 Bq source gives about 720 
count/s and (we estimate about 50\% efficiency for the detection of alpha
particles in the CsI detector, based on the fact that the Am$^{241}$ source
is mounted on a steel template). The expected number of 60 keV counts are
reduced by a fraction  $\alpha$ where $\alpha = \epsilon \Omega/(2 \pi) t $
where $\epsilon$ is the fraction of 60 keV X-rays emitted for each alpha
particle ($\sim$0.35), $\Omega$/(2 $\pi$) is the fractional solid angle
subtended by the Phoswich detector to the alpha source ($\sim$0.3) and
$t$ is the transmission of  CsI detector for the 60 keV X-rays ($\sim$0.35)
giving $\alpha$ to be 0.036 which agrees with the observed number of X-rays.

For the background estimation, the alpha-tag source was kept out of the
experimental set-up and necessary care was taken to avoid any X-ray photons 
from the alpha-tag source entering the PMT. The photon spectra were accumulated
for 100 seconds of exposure and collected using the MCA. This count spectrum 
are considered as the background spectrum. This is shown in Figure~\ref{fig4}.
After collecting the background spectra, the alpha-tag source was kept on the
Phoswich set-up and the corresponding spectra were taken. This  photon spectra
are shown in Figure~\ref{fig5}.

\begin{figure}
\centering
\includegraphics*[width=5.5cm,angle=-90]{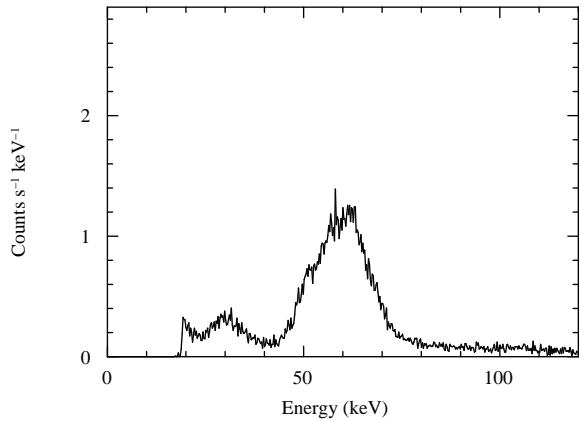}
\caption{The Phoswich detector output gated with Alpha-tag pulse.}
\label{fig6}
\end{figure}

Following these observations, the Alpha-tag detector was placed on the PMT
with the necessary experimental connections as shown above (Figure~\ref{fig3}).
The observations were repeated several times and the count spectrum for one 
such observation is shown in Figure~\ref{fig6}. We can see that the X-ray 
peak is enhanced considerably. Comparing the 80 -- 120 keV count rates in 
Figure~5 and Figure~6, we estimate that less than 10\% of background
X-rays are registered due to the tagging process. The total counts in
$<$ 80 keV region in Figure 6 is 27 s$^{-1}$. If we assume that the 
background in this region too is reduced by 90\%, we can estimate that
by the alpha-tagging method, more than 95\% of the 60 keV X-rays impinging
on the Phoswich detector are registered. This is somewhat higher than the 
50\% estimated earlier for the detection of alpha particles in the CsI 
detector. This discrepancy could be the result of a) the pallet of 
Am$^{241}$ could have a shape which enables a large fraction of alpha 
particles to impinge on the CsI crystal and/or b) the reduction of background
in the $<$ 80 keV region being higher than 10\% due to some threshold
effects. This experiment is done at somewhat higher background conditions 
(a total background of 177 s$^{-1}$). This alpha-tagged X-ray calibration 
source is proposed to be used in the Astrosat satellite where the onboard
background is expected to be only about 20 counts/s and hence negligible 
leakage is expected.

\section{Conclusions}

        We have developed a new alpha-tagged X-ray source for
space application which is rugged, cheap and easy to operate. 
Though the count rate is low (23 counts/s), it is sufficient to make 
onboard calibration with an integration time of about 100 s. Currently,
the leakage of background counts is about 10\%. This would be improved 
in a low background environment and better tuning of the gate-width for
the alpha detection. The alpha-tagging method selects only the
calibration X-rays along with negligible contribution from the 
background, thus enabling the calibration with a low count rate
source. Further, even if a small fraction of these X-rays enter the
detector without getting tagged (a few count per second), they
will have negligible impact on the overall background rates.

\section*{\bf Acknowledgment}

We thank Dr. K. Jahoda and an anonymous referee for useful suggestions
which improved the presentation of the paper. SN and RPSKK would like 
to acknowledge the hospitality provided by the Department of Astronomy 
and Astrophysics, Tata Institute of Fundamental Research during their 
visit to carry out the present work. The research work at Physical 
Research Laboratory  is funded by the Department of Space, Government 
of India.

\end{document}